\documentclass[twocolumn]{autart}    % Enable this line and disable the 
\usepackage{graphicx}          % Include this line if your 
\usepackage{dsfont} % For making double-stroke letters.
\usepackage[cmex10]{amsmath} % For among others the boldsymbol function.
\usepackage{varwidth} % This is for the box on the first page.
\usepackage{hyperref}

\newcommand{\ve}[1]{{\boldsymbol{#1}}} % For vectors in math mode.
\newcommand{\tr}{\mbox{tr}} % Short notation for the trace function.
\newcommand{\ex}{\mathds{E}} % Short notation for the expectation operator.
\newcommand{\va}{\mathds{V}} % Short notation for the variance operator.

\begin{document}
	
\journal{Automatica}
\volume{67}
\issue{May 2016}
\firstpage{216}
\lastpage{223}
\setcounter{page}{1}

\begin{frontmatter}
% Running title for regular papers but only if the title is over 5 words. Running title is not shown in output.
%\runtitle{Insert a suggested running title}
\title{Mean and variance of the LQG cost function}

\author[DCSC]{Hildo Bijl}\ead{h.j.bijl@tudelft.nl},
\author[DCSC]{Jan-Willem van Wingerden}\ead{j.w.vanwingerden@tudelft.nl},
\author[ITUU]{Thomas B. Sch\"on}\ead{thomas.schon@it.uu.se},
\author[DCSC]{Michel Verhaegen}\ead{m.verhaegen@tudelft.nl}

\address[DCSC]{Delft Center for Systems and Control, Delft University of Technology, The Netherlands}                                    
\address[ITUU]{Department of Information Technology, Uppsala University, Sweden}

\vspace{0.9cm}

\begin{center}
	\begin{varwidth}{14cm}
	\normalsize 
	\textbf{Please cite this version:}\\
	Hildo Bijl, Jan-Willem van Wingerden, Thomas B. Sch\"on, Michel Verhaegen. 
	Mean and variance of the LQG cost function. In \textit{Automatica}, Volume 67, May 2016, Pages 216-–223.\\
	\url{http://dx.doi.org/10.1016/j.automatica.2016.01.030}
	\end{varwidth}
\end{center}

\vspace{0.1cm}

% Five to ten keywords, chosen from the IFAC keyword list or with the help of the Automatica keyword wizard

% Abstract of not more than 200 words.
\begin{abstract}
Linear Quadratic Gaussian (LQG) systems are well-understood and methods to minimize the expected cost are readily available. Less is known about the statistical properties of the resulting cost function. The contribution of this paper is a set of analytic expressions for the mean and variance of the LQG cost function. These expressions are derived using two different methods, one using solutions to Lyapunov equations and the other using only matrix exponentials. Both the discounted and the non-discounted cost function are considered, as well as the finite-time and the infinite-time cost function. The derived expressions are successfully applied to an example system to reduce the probability of the cost exceeding a given threshold.
\end{abstract}

\end{frontmatter}

\section{Introduction}

The Linear-Quadratic-Gaussian (LQG) control paradigm is generally well-understood in literature. (See for instance \cite{LQBook,MFCBook,DMCSBook,StochasticControlBook}.) There are many methods available of calculating and minimizing the expected cost $\ex[J]$. However, much less is known about the resulting distribution of the cost function $J$. Yet in many cases (like in machine learning applications, in risk analysis and similar stochastic problems) knowledge of the full distribution of the cost function $J$, or at least knowledge of its variance $\va[J]$, is important. That is the focus of this paper. We derive analytical expressions for both the mean~$\ex[J]$ and the variance~$\va[J]$ of the cost function distribution for a variety of cases. The expressions for the variance~$\va[J]$ have not been published before, making that the main contribution of this paper.

The cost function~$J$ is usually defined as an integral over a squared non-zero-mean Gaussian process, turning its distribution into a generalized noncentral $\chi^2$~distribution. This distribution does not have a known Probability Density Function (PDF), although its properties have been studied before in literature, for instance in~\cite{MathematicalAnalysisOfRandomNoise,DistributionOfTheTimeAveragePower,LQGCostPDFApproximation}, and methods to approximate it are discussed in~\cite{QFRVBook,ChiSquareDistribution}. No expressions for the variance of the LQG system cost function are given though.

In LQG control most methods focus on the expected cost $\ex[J]$, but not all. For instance, Minimum Variance Control (MVC) (see \cite{StochasticControlBook}) minimizes the variance of the output $\ve{y}$, while Variance Constrained LQG (VCLQG) (see \cite{FuzzyWeightsForVCLQG,LQGControllerWithVarianceConstraints}) minimizes the cost function subject to bounds on the variance of the state $\ve{x}$ and/or the input $\ve{u}$. Alternatively, in Minimal Cost Variance (MCV) control (see~\cite{ControlUsingCostCumulants,StatisticalControlBook}) the mean cost~$\ex[J]$ is fixed through an equality constraint and the cost variance~$\va[J]$ (or alternatively the cost cumulant) is then minimized. However, expressions for the cost variance $\va[J]$ are still not given.

This paper is set up as follows. We present the problem formulation in Section~\ref{s:ProblemSetUp} and derive the expressions that solve this problem in Section~\ref{s:PropertiesOfLQGCost}, also making use of the appendices. Section~\ref{s:ApplicationToAnLQGSetUp} then shows how the equations can be applied to LQG systems, which is subsequently done in Section~\ref{s:NumericalEvaluation}. Finally, Section~\ref{s:ConclusionsAndRecommendations} contains the conclusions.

\section{Problem formulation} \label{s:ProblemSetUp}

We consider continuous linear systems subject to stochastic process noise. Formally, we write these as
\begin{equation}\label{eq:OfficialSystemDefinition}
d\ve{x}(t) = A\ve{x}(t) \, dt + d\ve{w}(t),
\end{equation}
where $\ve{w}(t)$ is a vector of Brownian motions. (Note that~\eqref{eq:OfficialSystemDefinition} is not an LQG system, because it is lacking input. The extension to LQG systems will be discussed in Section~\ref{s:ApplicationToAnLQGSetUp}.) As a result, $d\ve{w}(t)$ is a Gaussian random process with zero-mean and an (assumed constant) covariance of~$V \, dt$. Within the field of control (see for instance \cite{MFCBook}) this system is generally rewritten according to
\begin{equation}
\ve{\dot{x}}(t) = A\ve{x}(t) + \ve{v}(t), \label{eq:SystemDefinition}
\end{equation}
where~$\ve{v}(t)$ is zero-mean Gaussian white noise with intensity~$V$. That is, $\ex[\ve{v}(t) \ve{v}^T(\tau)] = V \delta(t-\tau)$, with $\delta(.)$ the Kronecker delta function. From a formal mathematical perspective this simplification is incorrect, because $\ve{v}(t)$ is not measurable with nonzero probability. However, since this notation is common in the control literature, and since it prevents us from having to evaluate the corresponding It\^o integrals, we will stick with it, although the reader is referred to~\cite{StochasticDEBook} for methods to properly deal with stochastic differential equations.

We assume that the initial state~$\ve{x}(0) = \ve{x}_0$ has a Gaussian distribution satisfying
\begin{equation}
\ve{\mu}_0 \equiv \ex[\ve{x}_0] \hspace{12pt} \mbox{and} \hspace{12pt} \Sigma_0 \equiv \ex[\ve{x}_0 \ve{x}_0^T].
\end{equation}
Note that the variance of~$\ve{x}_0$ is \textit{not}~$\Sigma_0$, but actually equals~\mbox{$\Sigma_0 - \ve{\mu}_0 \ve{\mu}_0^T$}. We will use two different cost functions in this paper: the infinite-time cost~$J$ and the finite-time cost~$J_T$, respectively defined as
\begin{align}
J & \equiv \int_0^\infty e^{2\alpha t} \ve{x}^T(t) Q \ve{x}(t) \, dt, \label{eq:InfiniteTimeCostFunction} \\
J_T & \equiv \int_0^T e^{2\alpha t} \ve{x}^T(t) Q \ve{x}(t) \, dt, \label{eq:FiniteTimeCostFunction}
\end{align}
where $Q$ is a user-defined symmetric weight matrix. The parameter~$\alpha$ can be positive or negative. If it is positive, it is known as the prescribed degree of stability (see \cite{LQBook} or \cite{DMCSBook}), while if it is negative (like in Reinforcement Learning applications) it is known as the discount exponent.

\section{Mean and variance of the LQG cost function} \label{s:PropertiesOfLQGCost}

In this section we derive expressions for $\ex[J]$, $\ex[J_T]$, $\va[J]$ and $\va[J_T]$. An overview of derived theorems, as well as the corresponding requirements, is shown in Table~\ref{t:TheoremOverview}.

\begin{table}[!t]
	\renewcommand{\arraystretch}{1.2}
	\caption{The theorems with which the mean and variance of $J$ and $J_T$ can be found, as well as the requirements for these theorems.}
	\label{t:TheoremOverview}
	\centering
	\begin{tabular}{|r||c|c|l|}
		\hline
		& \hspace{-3pt}If $\alpha \neq 0$\hspace{-3pt} & \hspace{-3pt}If $\alpha = 0$\hspace{-3pt} & \hspace{-3pt}Requirements \\
		\hline
		\hspace{-5pt} $\ex[J_T]$ \hspace{-5pt} & Th.~\ref{th:FiniteTimeCostMean} & Th.~\ref{th:FiniteTimeCostMeanWithAlphaZero} & \hspace{-3pt}$A$ and $A_\alpha$ Sylvester \\
		\hline
		\hspace{-5pt} $\ex[J]$ \hspace{-5pt} & 
		\multicolumn{2}{|c|}{Th.~\ref{th:InfiniteTimeCostMean}} & \hspace{-3pt}$\alpha < 0$ and $A_\alpha$ stable\hspace{-3pt} \\
		\hline
		\hspace{-5pt} $\va[J_T]$ \hspace{-5pt} & Th.~\ref{th:FiniteTimeCostVariance} & Th.~\ref{th:FiniteTimeCostVarianceWithAlphaZero} & \hspace{-3pt}$A_{-\alpha}$, $A$, $A_\alpha$ and $A_{2\alpha}$ Sylvester\hspace{-3pt} \\
		\hline
		\hspace{-5pt} $\ex[J]$ \hspace{-5pt} & 
		\multicolumn{2}{|c|}{Th.~\ref{th:InfiniteTimeCostVariance}} & \hspace{-3pt}$\alpha < 0$ and $A_\alpha$ stable\hspace{-3pt} \\
		\hline
	\end{tabular}
\end{table}

\subsection{Notation and terminology} \label{ss:NotationAndTerminology}

Concerning the evolution of the state, we define $\ve{\mu}(t) \equiv \ex[\ve{x}(t)]$, $\Sigma(t) \equiv \ex[\ve{x}(t)\ve{x}^T(t)]$ and $\Sigma(t_1,t_2) \equiv \ex[\ve{x}(t_1)\ve{x}^T(t_2)]$. These quantities can be found through the theorems of Appendix~\ref{s:StateEvolution}.

We define the matrices $A_\alpha \equiv A + \alpha I$ and similarly \mbox{$A_{k\alpha} \equiv A + k\alpha I$} for any number $k$. We also define $X^Q_{k\alpha}$ and $\bar{X}_{k\alpha}^Q$ to be the solutions of the Lyapunov equations
\begin{align}
A_{k\alpha} X_{k\alpha}^Q + X_{k\alpha}^Q A_{k\alpha}^T + Q & = 0, \\
A_{k\alpha}^T \bar{X}_{k\alpha}^Q + \bar{X}_{k\alpha}^Q A_{k\alpha} + Q & = 0.
\end{align}
We often have $\alpha = 0$. In this case $A_0$ equals $A$, and we similarly shorten $X^Q_0$ to $X^Q$. The structure inherent in the Lyapunov equation induces interesting properties in its solutions $X_{k\alpha}^Q$, which are outlined in Appendix~\ref{s:LyapunovTheorems}.

We define the time-dependent solution $X_{k\alpha}^Q(t_1,t_2)$ as
\begin{equation}
X^Q_{k\alpha}(t_1,t_2) = \int_{t_1}^{t_2} e^{A_{k\alpha} t} Q e^{A_{k\alpha}^T t} \, dt. \label{eq:XQTDefinition}
\end{equation}
This integral can be calculated efficiently by solving a Lyapunov equation. (See Theorem~\ref{th:FiniteIntegralToLyapunov}.) Often it happens that the lower limit $t_1$ of $X_{k\alpha}^Q(t_1,t_2)$ equals zero. To simplify notation, we then write $X_{k\alpha}^Q(t) \equiv X_{k\alpha}^Q(0,t)$. Another integral solution $\tilde{X}_{k_1\alpha,k_2\alpha}^Q(T)$ is defined as
\begin{equation}
\tilde{X}_{k_1\alpha,k_2\alpha}^Q(T) \equiv \int_0^T e^{A_{k_1\alpha} (T-t)} Q e^{A_{k_2\alpha} t} \, dt.
\end{equation}
This quantity can be calculated (see \cite{MatrixExponentials}) through
\begin{equation}
\tilde{X}_{\alpha_1,\alpha_2}^Q(T) = \begin{bmatrix}
I & 0
\end{bmatrix} \exp\left(\begin{bmatrix}
A_{\alpha_1} & Q \\
0 & A_{\alpha_2}
\end{bmatrix} T\right) \begin{bmatrix}
0 \\
I
\end{bmatrix}. \label{eq:CalculatingXTilde}
\end{equation}
Considering terminology, we say that a matrix $A$ is \textit{stable} (Hurwitz) if and only if it has no eigenvalue $\lambda_i$ with a real part equal to or larger than zero. Similarly, we say that a matrix $A$ is \textit{Sylvester} if and only if it has no two eigenvalues $\lambda_i$ and $\lambda_j$ (with possibly $i = j$) satisfying $\lambda_i = -\lambda_j$. This latter definition is new in literature, but to the best of our knowledge, no term for this matrix property has been defined earlier.

\subsection{The expected cost}

We now examine the expected costs $\ex[J]$ and $\ex[J_T]$. Expressions for these costs are already known for various special cases. (See for instance \cite{DMCSBook,StochasticControlBook}.) To provide a complete overview of the subject, we have included expressions which are as general as possible.

\begin{thm}\label{th:FiniteTimeCostMean}
Consider system~\eqref{eq:SystemDefinition}. Assume that $\alpha \neq 0$ and that $A$ and $A_\alpha$ are both Sylvester. The expected value $\ex[J_T]$ of the finite-time cost $J_T$~\eqref{eq:FiniteTimeCostFunction} then equals
\begin{equation}
\tr\left(\hspace{-2pt}\left(\hspace{-2pt}\Sigma_0 - e^{2\alpha T} \Sigma(T) + \left(1 - e^{2\alpha T}\right)\hspace{-2pt}\left(\hspace{-1pt}\frac{-V}{2\alpha}\hspace{-1pt}\right)\hspace{-2pt}\right) \bar{X}_\alpha^Q\hspace{-2pt}\right). \label{eq:FiniteTimeCostMean}
\end{equation}
\end{thm}
\begin{pf}
From~\eqref{eq:FiniteTimeCostFunction} follows directly that
\begin{equation}
\ex[J_T] = \tr\left(\int_0^T e^{2\alpha t} \Sigma(t) \, dt \, Q\right) = \tr\left(Y(T)Q\right), \label{eq:EJTIntermediateExpression}
\end{equation}
where $Y(T)$ is defined as the above integral. To find it, we multiply~\eqref{eq:StateProcessDerivative} by $e^{2\alpha t}$ and integrate it to get
\begin{equation}
\int_0^T \hspace{-4pt} e^{2\alpha t} \dot{\Sigma}(t) \, dt = A Y(T) + Y(T) A^T + \int_0^T \hspace{-4pt} e^{2\alpha t} V \, dt.
\end{equation}
The left part, through integration by parts, must equal
\begin{equation}
\int_0^T e^{2\alpha t} \dot{\Sigma}(t) \, dt = \left(e^{2\alpha T} \Sigma(T) - \Sigma_0\right) - 2\alpha Y(T).
\end{equation}
As a result, $Y(T)$ must satisfy the Lyapunov equation
\begin{equation}
A_\alpha Y(T) + Y(T) A_\alpha^T + \left(\hspace{-2pt}\frac{e^{2\alpha T} - 1}{2\alpha} V + \Sigma_0 - e^{2\alpha T} \Sigma(T)\hspace{-2pt}\right) \hspace{-2pt}=\hspace{-1pt} 0.
\end{equation}
Using Theorem~\ref{th:AddingLyapunovSolutions}, we can now write $Y(T)$ as
\begin{equation}
Y(T) = \frac{e^{2\alpha T} - 1}{2\alpha}X_\alpha^V + X_\alpha^{\Sigma_0} - e^{2\alpha T} X_\alpha^{\Sigma(T)}.
\end{equation}
Combining this with~\eqref{eq:EJTIntermediateExpression} and applying Theorem~\ref{th:InterchangeLyapunovSolutions} (with $F = G = I$) completes the proof.
\end{pf}

\begin{thm}\label{th:InfiniteTimeCostMean}
Consider system~\eqref{eq:SystemDefinition}. Assume that $\alpha < 0$ and that $A_\alpha$ is stable. The expected value $\ex[J]$ of the infinite-time cost $J$~\eqref{eq:InfiniteTimeCostFunction} is then given by
\begin{equation}
\ex[J] = \tr\left(\left(\Sigma_0 - \frac{V}{2\alpha}\right) \bar{X}_\alpha^Q\right). \label{eq:InfiniteTimeCostMean}
\end{equation}
\end{thm}
\begin{pf}
If we examine~\eqref{eq:FiniteTimeCostMean} in the limit as $T \rightarrow \infty$, then this theorem directly follows. After all, Theorem~\ref{th:PropertiesOfStateProcess} implies that, for stable $A_\alpha$, $e^{2\alpha T} \Sigma(T) \rightarrow 0$ as $T \rightarrow \infty$.
\end{pf}

\begin{thm}\label{th:FiniteTimeCostMeanWithAlphaZero}
Consider system~\eqref{eq:SystemDefinition}. Assume that $\alpha = 0$ and that $A$ is Sylvester. The expected value $\ex[J_T]$ of the finite-time cost $J_T$~\eqref{eq:FiniteTimeCostFunction} is then given by
\begin{equation}
\ex[J_T] = \tr\left(\left(\Sigma_0 - \Sigma(T) + TV\right) \bar{X}^Q\right). \label{eq:FiniteTimeCostMeanWithAlphaZero}
\end{equation}
\end{thm}
\begin{pf}
If we consider~\eqref{eq:FiniteTimeCostMean} from Theorem~\ref{th:FiniteTimeCostMean} as $\alpha \rightarrow 0$, then this theorem directly follows. After all, we know from l'H\^opital's rule that $\lim_{\alpha \rightarrow 0} \frac{1 - e^{2\alpha T}}{2\alpha} = -T$.
\end{pf}

\subsection{The cost variance}

Next, we derive expressions for the variances $\va[J]$ and $\va[J_T]$. These expressions are new and as such are our main contribution. If we define $\Delta = \Sigma_0 - X^V$, then $\va[J_T]$ and $\va[J]$ can be found through the following theorems.

\begin{thm}\label{th:FiniteTimeCostVariance}
Consider system~\eqref{eq:SystemDefinition}. Assume that $\alpha \neq 0$ and that $A_{-\alpha}$, $A$, $A_\alpha$ and $A_{2\alpha}$ are Sylvester. The variance $\va[J_T]$ of the finite-time cost $J_T$~\eqref{eq:FiniteTimeCostFunction} is then given by
\begin{align}
\va[J_T] & = 2\tr\left((\Delta \bar{X}_\alpha^Q(T))^2\right) - 2\left(\ve{\mu}_0^T \bar{X}_\alpha^Q(T) \ve{\mu}_0\right)^2 \nonumber \\
& \hspace{18pt} + 4\tr\bigg(X^V Q \bigg(X^V \frac{e^{4\alpha T} \bar{X}_{-\alpha}^Q(T) - \bar{X}_\alpha^Q(T)}{4\alpha} \nonumber \\
& \hspace{18pt} + 2X_{2\alpha}^\Delta \bar{X}_\alpha^Q(T) - 2\tilde{X}_{3\alpha,\alpha}^{X_{2\alpha}^\Delta e^{A_\alpha^T T} Q}(T)\bigg)\bigg). \label{eq:FiniteTimeCostVariance}
\end{align}
\end{thm}
\begin{pf}
We will start our proof by evaluating $\ex[J^2]$. If we write $\ve{x}(t_1)$ as $\ve{x}_1$ and $\ve{x}(t_2)$ as $\ve{x}_2$, then we have
\begin{equation}
\ex[J^2] \hspace{-2pt}=\hspace{-2pt} \ex\left[\hspace{-1pt}\int_0^T \hspace{-6pt} \int_0^T \hspace{-8pt} e^{2\alpha (t_1 + t_2)} \ve{x}_1^T Q \ve{x}_1 \ve{x}_2^T Q \ve{x}_2 \, dt_2 \, dt_1\hspace{-1pt}\right]\hspace{-2pt}.
\end{equation}
Taking the trace and applying Theorem~\ref{th:ExpectationOfDifferentGaussianPowerFour} gives us
\begin{align}
\ex[J^2] & = \int_0^T \hspace{-6pt} \int_0^T \hspace{-4pt} \Big( \tr\left(e^{2\alpha t_1}\Sigma(t_1) Q\right) \tr\left(e^{2\alpha t_2}\Sigma(t_2) Q\right) \nonumber \\
& \hspace{20pt} + 2\tr\left(e^{2\alpha(t_1+t_2)}\Sigma(t_2,t_1) Q \Sigma(t_1,t_2) Q\right) \nonumber \\
& \hspace{20pt} - 2e^{2\alpha(t_1+t_2)}\ve{\mu}_1^T Q \ve{\mu}_1 \ve{\mu}_2^T Q \ve{\mu}_2 \Big) \, dt_2 \, dt_1, \label{eq:WorkedOutExpressionEJ2}
\end{align}
where $\ve{\mu}_1$ equals $\ex[\ve{x}(t_1)] = e^{At_1} \ve{\mu}_0$ (see Theorem~\ref{th:PropertiesOfStateProcess}) and similarly for $\ve{\mu}_2$. There are three terms in the above equation. We will denote them by $T_1$, $T_2$ and $T_3$, respectively. The first term $T_1$ directly equals $\ex[J]^2$ (see Theorem~\ref{th:FiniteTimeCostMean}). This is convenient, because $\va[J] = \ex[J^2] - \ex[J]^2$, which means that $\va[J]$ equals the remaining two terms $T_2 + T_3$.

The third term $T_3$ is, according to definition~\eqref{eq:XQTDefinition}, equal to
\begin{align}
T_3 & = -2\left(\int_0^T \hspace{-6pt} e^{2\alpha t} \ve{\mu}_0^T e^{A^T t} Q e^{A t} \ve{\mu}_0 \, dt\right)^2 \hspace{-5pt} \nonumber \\
& = -2\left(\ve{\mu}_0^T \bar{X}_\alpha^Q(T) \ve{\mu}_0\right)^2, \label{eq:T3Expression}
\end{align}
where $\bar{X}_\alpha^Q(T)$ can be evaluated through Theorem~\ref{th:FiniteIntegralToLyapunov}. That leaves $T_2$. To find it, we first have to adjust the integrals. We note that $T_2$ is symmetric with respect to $t_1$ and $t_2$. That is, if we would interchange $t_1$ and $t_2$, the integrand would be the same. As a result, we do not have to integrate over all values of $t_1$ and $t_2$. We can also only consider all cases where $t_1 < t_2$, integrate over this area, and then multiply the final result by $2$. This gives us
\begin{equation}
T_2 = 4\tr\left(\int_0^T \hspace{-6pt} \int_{t_1}^T \hspace{-6pt} e^{2\alpha(t_1+t_2)}\Sigma(t_2,t_1) Q \Sigma(t_1,t_2) Q \, dt_2 \, dt_1\right). \label{eq:TrickOfUsingSymmetry}
\end{equation}
Now, with $t_1 < t_2$, we can apply Theorem~\ref{th:MultipleTimeStateProcess} to substitute for $\Sigma(t_1,t_2)$. If we subsequently expand the brackets, and use the fact that $X^V$ and hence also $\Delta$ is symmetric (see Theorem~\ref{th:SymmetricLyapunovSolution}), then the above term turns into
\begin{align}
T_2 & = 4\tr\bigg(\int_0^T \hspace{-6pt} \int_{t_1}^T \hspace{-6pt} e^{2\alpha(t_1+t_2)} \bigg(e^{A t_2} \Delta e^{A^T t_1} Q e^{A t_1} \Delta e^{A^T t_2} Q \nonumber \\
& \hspace{16pt} + e^{A(t_2-t_1)} X^V Q X^V e^{A^T (t_2-t_1)} Q \nonumber \\
& \hspace{16pt} + 2 e^{A (t_2-t_1)} X^V Q e^{At_1} \Delta e^{A^T t_2} Q \bigg) \, dt_2 \, dt_1\bigg).
\end{align}
This expression again has three terms. We call them $T_{2,1}$, $T_{2,2}$ and $T_{2,3}$, respectively. First we find $T_{2,1}$. We can again note that the integrand is symmetric with respect to $t_1$ and $t_2$, meaning we can apply the opposite trick of the one we applied at~\eqref{eq:TrickOfUsingSymmetry}. This gives us
\begin{align}
& T_{2,1} = 2\tr\left(\int_0^T \hspace{-6pt} \int_0^T \hspace{-6pt} e^{A_\alpha t_2} \Delta e^{A_\alpha^T t_1} Q e^{A_\alpha t_1} \Delta e^{A_\alpha^T t_2} Q \, dt_2 \, dt_1\right) \nonumber \\
& \hspace{-6pt} = \hspace{-1pt} 2\tr\hspace{-2pt}\left(\hspace{-4pt}\left.\left(\hspace{-1pt}\int_0^T \hspace{-8pt} \Delta e^{A_\alpha^T t} Q e^{A_\alpha t} \, dt\hspace{-1pt}\right)\hspace{-5pt}\right.^2 \right) \hspace{-2pt} = \hspace{-1pt} 2\tr\hspace{-2pt}\left((\Delta \bar{X}_\alpha^Q(T))^2\right). \label{eq:T21Expression}
\end{align}
The next term, $T_{2,2}$, is not symmetric in $t_1$ and $t_2$. To bring both integration bounds back to zero, we now substitute $t_2$ for $t_2 + t_1$. Subsequently interchanging the integrals results in
\begin{align}
T_{2,2} & \hspace{-1pt}=\hspace{-1pt} 4\tr\bigg(\hspace{-1pt}\int_0^T \hspace{-6pt} \int_0^{T-t_1} \hspace{-17pt} e^{2\alpha(2t_1 + t_2)} e^{At_2} X^V Q X^V e^{A^T t_2} Q \, dt_2 \, dt_1\hspace{-2pt}\bigg) \nonumber \\
& \hspace{-1pt}=\hspace{-1pt} 4\tr\left(\hspace{-1pt}\int_0^T \hspace{-5pt} \left(\hspace{-1pt}\int_0^{T-t_2} \hspace{-17pt} e^{4\alpha t_1} dt_1\hspace{-2pt}\right)\hspace{-2pt} e^{A_\alpha ^T t_2} Q e^{A_\alpha t_2} X^V Q X^V dt_2\right) \nonumber \\
& \hspace{-1pt}=\hspace{-1pt} 4\tr\left(\hspace{-2pt}\frac{e^{4\alpha T} \bar{X}_{-\alpha}^Q(T) - \bar{X}_\alpha^Q(T)}{4\alpha} X^V Q X^V\hspace{-2pt}\right).
\end{align}
That leaves $T_{2,3}$, which is the most involved term. We can apply the same substitution and interchanging of integrals to find that $T_{2,3}$ equals
\begin{align}
& 8\tr\hspace{-2pt}\left(\hspace{-1pt}\int_0^T \hspace{-6pt} \int_0^{T-t_2} \hspace{-20pt} e^{2\alpha(2t_1+t_2)} e^{A t_2} X^V Q e^{A t_1} \Delta e^{A^T (t_2 + t_1)} Q \, dt_1 \, dt_2\hspace{-2pt}\right) \nonumber \\
& \hspace{-4pt}=\hspace{-2pt} 8\tr\hspace{-2pt}\left(\hspace{-1pt}\int_0^T \hspace{-9pt} X^V Q X_{2\alpha}^\Delta(T\hspace{-3pt}-\hspace{-2pt}t_2) e^{A_\alpha^T t_2} Q e^{A_\alpha t_2} \, dt_2 \hspace{-2pt}\right) \hspace{-2pt} \hspace{-1pt}=\hspace{-2pt} T_{2,3}.
\end{align}
Expanding $X_{2\alpha}^\Delta(T-t_2)$ using Theorem~\ref{th:FiniteIntegralToLyapunov} turns this into
\begin{align}
\hspace{-5pt} T_{2,3} & = 8\tr\hspace{-1pt}\Bigg(X^V Q \Bigg(X_{2\alpha}^\Delta \int_0^T e^{A_\alpha^T t_2} Q e^{A_\alpha t_2} \, dt_2 \nonumber \\
& \hspace{12pt} - \int_0^T e^{A_{3\alpha}(T - t_2)} X_{2\alpha}^\Delta e^{A_\alpha^T T} Q e^{A_\alpha t_2} \, dt_2\Bigg)\Bigg) \\
& = 8\tr\hspace{-1pt}\left(X^V Q \left(X_{2\alpha}^\Delta \bar{X}_\alpha^Q(T) - \tilde{X}_{3\alpha,\alpha}^{X_{2\alpha}^\Delta e^{A_\alpha^T T} Q}(T)\right)\right), \nonumber
\end{align}
where the final term $\tilde{X}_{3\alpha,\alpha}^{X_{2\alpha}^\Delta e^{A_\alpha^T T} Q}(T)$ can be found through~\eqref{eq:CalculatingXTilde}. If we now merge all terms together, we find the result which we wanted to prove.
\end{pf}

\begin{thm}\label{th:InfiniteTimeCostVariance}
Consider system~\eqref{eq:SystemDefinition}. Assume that $\alpha < 0$ and that $A_\alpha$ is stable. The variance $\va[J]$ of the infinite-time cost $J$~\eqref{eq:InfiniteTimeCostFunction} is then given by
\begin{align}
\va[J] & = 2\tr\left((\Sigma_0 \bar{X}_\alpha^Q)^2\right) - 2\left(\ve{\mu}_0^T \bar{X}_\alpha^Q \ve{\mu}_0\right)^2 \nonumber \\
& \hspace{12pt} + 4\tr\left(\left(X_{2\alpha}^{\Sigma_0} - \frac{X_{2\alpha}^V}{4\alpha}\right) \bar{X}_\alpha^Q V \bar{X}_\alpha^Q\right). \label{eq:InfiniteTimeCostVariance}
\end{align}
\end{thm}
\begin{pf}
As $T \rightarrow \infty$, $e^{A_\alpha^T T}$ and $e^{4\alpha T}$ become zero, $\bar{X}_\alpha^Q(T)$ becomes $\bar{X}_\alpha^Q$ and hence~\eqref{eq:FiniteTimeCostVariance} reduces to
\begin{align}
\va[J] & = 2\tr\left((\Delta \bar{X}_\alpha^Q)^2\right) - 2\left(\ve{\mu}_0^T \bar{X}_\alpha^Q \ve{\mu}_0\right)^2 \label{eq:InfiniteTimeCostVarianceReduction} \\
& \hspace{10pt} + 4\tr\left(\bar{X}_\alpha^Q X^V Q \left(2X_{2\alpha}^\Delta - \frac{X^V}{4\alpha}\right)\right). \nonumber
\end{align}
Through an excessive amount of elementary rewritings, using both $Q = -A_\alpha^T \bar{X}_\alpha^Q - \bar{X}_\alpha^Q A_\alpha$ and Theorem~\ref{th:DifferenceBetweenLyapunovSolutions}, the above can be rewritten to~\eqref{eq:InfiniteTimeCostVariance}, which is a slightly more elegant version of the above expression.
\end{pf}

\begin{thm}\label{th:FiniteTimeCostVarianceWithAlphaZero}
Consider system~\eqref{eq:SystemDefinition}. Assume that $\alpha = 0$ and that $A$ is Sylvester. The variance $\va[J_T]$ of the finite-time cost $J_T$~\eqref{eq:FiniteTimeCostFunction} is then given by
\begin{align}
\va[J_T] & = 2\tr\left((\Delta \bar{X}^Q(T))^2\right) - 2\left(\ve{\mu}_0^T \bar{X}^Q(T) \ve{\mu}_0\right)^2 \nonumber \\
& \hspace{18pt} + 4\tr\bigg(X^V Q \bigg(X^V \left(T\bar{X}^Q - X^{X^Q}(T)\right) \nonumber \\
& \hspace{18pt} + 2X^\Delta \bar{X}^Q(T) - 2\tilde{X}^{X^\Delta e^{A^T T} Q}(T)\bigg)\bigg). \label{eq:FiniteTimeCostVarianceWithAlphaZero}
\end{align}
\end{thm}
\begin{pf}
We can evaluate~\eqref{eq:FiniteTimeCostVariance} from Theorem~\ref{th:FiniteTimeCostVariance} as $\alpha \rightarrow 0$. While doing so, we may use the relation
\begin{equation}
\bar{X}_\alpha^{\bar{X}_{-\alpha}^Q}\hspace{-1pt}(T) \hspace{-2pt}=\hspace{-2pt} \frac{\bar{X}_\alpha^Q(T) \hspace{-2pt}-\hspace{-2pt} e^{4\alpha T} \bar{X}_{-\alpha}^Q(T)}{4\alpha} \hspace{-2pt}+\hspace{-2pt} \frac{e^{4\alpha T} \hspace{-3pt}-\hspace{-2pt} 1}{4\alpha} \bar{X}_{-\alpha}^Q,
\end{equation}
which follows from combining Theorems~\ref{th:FiniteIntegralToLyapunov} and~\ref{th:DifferenceBetweenLyapunovSolutions}. From this, we find through application of l'H\^opital's rule that
\begin{equation}
\lim_{\alpha \rightarrow 0} \frac{e^{4\alpha T} \bar{X}_{-\alpha}^Q(T) \hspace{-1pt}-\hspace{-1pt} \bar{X}_\alpha^Q(T)}{4\alpha} \hspace{-1pt}=\hspace{-1pt} T\bar{X}^Q \hspace{-1pt}-\hspace{-1pt} X^{X^Q}(T).
\end{equation}
By using the above relation, the theorem directly follows.
\end{pf}

\subsection{Finding $\ex[J_T]$ and $\va[J_T]$ using matrix exponentials}

The method of using Lyapunov solutions to find $\ex[J_T]$ and $\va[J_T]$ has a significant downside: if $A$ or $A_\alpha$ is not Sylvester, the theorems do not hold. By solving integrals using matrix exponentials, according to the methods described in~\cite{MatrixExponentials}, we can work around that problem.

\begin{thm}\label{th:MatrixExponentialsForCostMeanAndVariance}
If we define the matrix $C$ as
\begin{equation}
C = \begin{bmatrix}
-A_{2\alpha}^T & Q & 0 & 0 & 0 \\
0 & A & V & 0 & 0 \\
0 & 0 & -A^T & Q & 0 \\
0 & 0 & 0 & A_{2\alpha} & V \\
0 & 0 & 0 & 0 & -A_{-2\alpha}^T
\end{bmatrix},
\end{equation}
and write $e^{CT}$ as
\begin{equation}
e^{CT} = \begin{bmatrix}
C_{11}^e & \cdots & C_{15}^e \\
\vdots & \ddots & \vdots \\
C_{51}^e & \cdots & C_{55}^e
\end{bmatrix},
\end{equation}
then we can find $\ex[J_T]$ and $\va[J_T]$ through
\begin{align}
\ex[J_T] & = \tr\left((C_{44}^e)^T \left(C_{12}^e \Sigma_0 + C_{13}^e\right)\right), \label{eq:FiniteTimeMatrixExponentialCostMean} \\
\va[J_T] & = 2\tr\Big(\hspace{-3pt}\left((C_{44}^e)^T \left(C_{12}^e \Sigma_0 + C_{13}^e\right)\right)^2 \label{eq:FiniteTimeMatrixExponentialCostVariance} \\
& \hspace{-10pt} - 2(C_{44}^e)^T (C_{14}^e \Sigma_0 + C_{15}^e)\Big) - 2\left(\ve{\mu}_0^T (C_{44}^e)^T C_{12}^e \ve{\mu}_0\right)^2. \nonumber
\end{align}
\end{thm}
\begin{pf}
We first prove the expression for $\ex[J_T]$. If we insert~\eqref{eq:SigmaTIntermediateExpression} into~\eqref{eq:EJTIntermediateExpression}, we find that
\begin{align}
\ex[J_T] & = \tr\bigg(\int_0^T e^{2\alpha t} e^{A t} \Sigma_0 e^{A^T t} Q \, dt \nonumber \\
& \hspace{10pt} + \int_0^T \int_0^t e^{2\alpha t} e^{A (t-s)} V e^{A^T (t-s)} Q \, ds \, dt\bigg).
\end{align}
We know from~\cite{MatrixExponentials} that
\begin{align}
C_{44}^e & = e^{A_{2\alpha} T}, \\
C_{12}^e & = \int_0^T e^{-A_{2\alpha}^T (T-t)} Q e^{A t} \, dt, \\
C_{13}^e & = \int_0^T \int_0^t e^{-A_{2\alpha}^T (T-t)} Q e^{A (t-s)} V e^{-A^T s} \, ds \, dt.
\end{align}
From this~\eqref{eq:FiniteTimeMatrixExponentialCostMean} directly follows. Proving the expression for $\va[J_T]$ is done similarly, but with more bookkeeping. First of all, $C_{14}^e$ equals (see~\cite{MatrixExponentials})
\begin{equation}
\int_0^T \hspace{-6pt} \int_0^t \hspace{-4pt} \int_0^s \hspace{-6pt} e^{-A_{2\alpha}^T (T-t)} Q e^{A (t-s)} V e^{-A^T (s-r)} Q e^{A_{2\alpha} r} \, dr \, ds \, dt,
\end{equation}
with a similar expression for $C_{15}^e$. Next, we will find the terms $T_3$ (see~\eqref{eq:T3Expression}) and $T_2$ (see~\eqref{eq:TrickOfUsingSymmetry}), which together equal $\va[J_T]$. We can directly see from~\eqref{eq:T3Expression} that $T_3$ equals
\begin{equation}
T_3 = -2\left(\ve{\mu}_0^T (C_{44}^e)^T C_{12}^e \ve{\mu}_0\right)^2.
\end{equation}
Then we consider $T_2$ from~\eqref{eq:TrickOfUsingSymmetry}. Instead of applying~\eqref{eq:SigmaTIntermediateExpression}, we now use
\begin{equation}
\hspace{-1pt}\Sigma(t_1,t_2) \hspace{-1pt}=\hspace{-1pt} e^{A t_1} \Sigma_0 e^{A^T t_2} \hspace{-0pt}+\hspace{-2pt} \int_0^{\min(t_1,t_2)} \hspace{-34pt} e^{A (t_1 - s)} V e^{A^T (t_2 - s)} ds, \hspace{-2pt}
\end{equation}
which is derived in an identical way. For ease of notation, we write $\Sigma(t_1,t_2) = \Sigma_a + \Sigma_b$, with $\Sigma_a$ and $\Sigma_b$ the two parts in the above expression. Inserting $\Sigma(t_1,t_2)$ into~\eqref{eq:TrickOfUsingSymmetry} then gives
\begin{align}
T_2 & = 2\tr\bigg(\int_0^T \int_0^T e^{2\alpha(t_1+t_2)} \bigg(\Sigma_a^T Q \Sigma_a Q \nonumber \\
& \hspace{20pt} + 2\Sigma_a^T Q \Sigma_b Q + \Sigma_b^T Q \Sigma_b Q\bigg) \, dt_2 \, dt_1 \bigg).
\end{align}
The first term $T_{2,aa}$ here equals
\begin{align}
& 2\tr\left(\hspace{-1pt}\int_0^T \hspace{-6pt} \int_0^T \hspace{-6pt} e^{2\alpha(t_1+t_2)} e^{A t_2} \Sigma_0 e^{A^T t_1} Q e^{A t_1} \Sigma_0 e^{A^T t_2} \, dt_2 \, dt_1\hspace{-2pt}\right) \nonumber \\
& \hspace{20pt} = 2\tr\left(\left(\int_0^T e^{2\alpha t} e^{A^T t} Q e^{A t} \Sigma_0 \, dt\right)^2\right) \nonumber \\
& \hspace{20pt} = 2\tr\left(\left((C_{44}^e)^T C_{12}^e \Sigma_0\right)^2\right) = T_{2,aa}.
\end{align}
The second term $T_{2,ab}$ is given by
\begin{align}
T_{2,ab} & = 4\tr\bigg(\int_0^T \int_0^T \int_0^{\min(t_1,t_2)} \hspace{-24pt} e^{2\alpha(t_1+t_2)} e^{A t_2} \Sigma_0 e^{A^T t_1} Q \nonumber \\
& \hspace{32pt} e^{A (t_1 - s)} V e^{A^T (t_2 - s)} Q \, ds \, dt_2 \, dt_1\bigg).
\end{align}
We want the integration order to be $dt_2 \, ds \, dt_1$. If we note that the integration area is described by $0 \leq s \leq (t_1,t_2) \leq T$, we can reorder the integrals. That is,
\begin{align}
T_{2,ab} & \hspace{-2pt}=\hspace{-2pt} 4\tr\bigg(\hspace{-2pt} \int_0^T \hspace{-6pt} \int_0^{t_1} \hspace{-6pt} \int_s^T \hspace{-6pt} \ldots dt_2 \, ds \, dt_1 \hspace{-2pt} \bigg) \\
& \hspace{-2pt}=\hspace{-2pt} 4\tr\bigg(\hspace{-2pt} \int_0^T \hspace{-6pt} \int_0^{t_1} \hspace{-6pt} \int_0^T \hspace{-7pt} \ldots dt_2 \, ds \, dt_1 \hspace{-2pt}-\hspace{-2pt} \int_0^T \hspace{-6pt} \int_0^{t_1} \hspace{-6pt} \int_0^s \hspace{-7pt} \ldots dt_2 \, ds \, dt_1 \hspace{-2pt} \bigg). \nonumber
\end{align}
We now have two integrals, but we can solve both. If we split up the first one and rewrite the second one, we get
\begin{align}
\hspace{-2pt}T_{2,ab} & \hspace{-2pt}=\hspace{-2pt} 4\tr\bigg(\hspace{-2pt}\bigg(\hspace{-1pt}\int_0^T \hspace{-6pt} \int_0^{t_1} \hspace{-6pt} e^{2\alpha t_1} e^{A^T t_1} Q e^{A (t_1 - s)} V e^{-A^T s} \, ds \, dt_1\hspace{-2pt}\bigg) \nonumber \\
& \hspace{6pt} \bigg(\hspace{-2pt}\int_0^T \hspace{-6pt} e^{2\alpha t_2} e^{A^T t_2} Q e^{A t_2} \, dt_2\hspace{-2pt}\bigg) \Sigma_0 \hspace{-2pt}-\hspace{-2pt} \int_0^T \hspace{-6pt} \int_0^{t_1} \hspace{-6pt} \int_0^s \hspace{-6pt} e^{2\alpha(t_1+t_2)} \nonumber \\
& \hspace{18pt} e^{A^T t_1} Q e^{A (t_1 - s)} V e^{A^T (t_2 - s)} Q e^{A t_2} \Sigma_0 \, dt_2 \, ds \, dt_1\hspace{-2pt}\bigg) \nonumber \\
& \hspace{-2pt}=\hspace{-2pt} 4\tr\hspace{-2pt}\left(\hspace{-1pt}(C_{44}^e)\hspace{-1pt}^T\hspace{-1pt} C_{13}^e (C_{44}^e)\hspace{-1pt}^T\hspace{-1pt} C_{12}^e \Sigma_0 \hspace{-2pt}-\hspace{-2pt} (C_{44}^e)\hspace{-1pt}^T\hspace{-1pt} C_{14}^e \Sigma_0\hspace{-1pt}\right)\hspace{-2pt}.
\end{align}
Finally there is $T_{2,bb}$. We first concern ourselves with the integration order and limits. By rearranging integrals, and by using the symmetry between $t_1$ and $t_2$ as well as between $s_1$ and $s_2$, we can find that
\begin{align}
T_{2,bb} & = 2\tr\bigg(\hspace{-1pt}\int_0^T \hspace{-6pt} \int_0^T \hspace{-6pt} \int_0^{\min(t_1,t_2)} \hspace{-6pt} \int_0^{\min(t_1,t_2)} \hspace{-16pt} \ldots \, ds_2 \, ds_1 \, dt_2 \, dt_1\hspace{-2pt}\bigg) \nonumber \\
& = 2\tr\bigg(\hspace{-1pt}\int_0^T \hspace{-6pt} \int_0^{t_2} \hspace{-6pt} \int_0^T \hspace{-6pt} \int_0^{t_1} \hspace{-6pt} \ldots \, ds_1 \, dt_2 \, ds_2 \, dt_1 \nonumber \\
& \hspace{32pt} - 2\int_0^T \hspace{-6pt} \int_0^{t_2} \hspace{-6pt} \int_0^{s_1} \hspace{-6pt} \int_0^{t_1} \hspace{-6pt} \ldots \, ds_1 \, dt_2 \, ds_2 \, dt_1\hspace{-2pt}\bigg).
\end{align}
After inserting the integrand, we can rewrite this to
\begin{align}
T_{2,bb} & = 2\tr\bigg(\bigg(\int_0^T \hspace{-6pt} \int_0^t e^{2\alpha t} e^{A^T t} Q e^{A(t-s)} V e^{-A^T s} \, ds \, dt\bigg)^2 \nonumber \\
& \hspace{10pt} - 2\int_0^T \hspace{-6pt} \int_0^{t_2} \hspace{-6pt} \int_0^{s_1} \hspace{-6pt} \int_0^{t_1} \hspace{-6pt} e^{2\alpha(t_1+t_2)} e^{A^T (t_1 - s_1)} Q e^{A (t_1 - s_2)} V \nonumber \\
& \hspace{56pt} e^{A^T (t_2 - s_2)} Q e^{A (t_2 - s_1)} V \, ds_1 \, dt_2 \, ds_2 \, dt_1\bigg) \nonumber \\
& = 2\tr\left(\hspace{-1pt}\left((C_{44}^e)^T C_{13}^e\right)^2 \hspace{-2pt}-\hspace{-2pt} 2 (C_{44}^e)^T C_{15}^e\hspace{-1pt}\right).
\end{align}
By combining all the results, we wind up with~\eqref{eq:FiniteTimeMatrixExponentialCostVariance}.
\end{pf}

So now we have two methods of finding $\ex[J_T]$ and $\va[J_T]$. But which one is better? This mainly depends on the time $T$. Our experiments have shown that, for small times $T$, using matrix exponentials results in a better numerical accuracy than using Lyapunov solutions, but for large $T$ the situation is exactly the opposite, and the numerical accuracy of the matrix exponential method quickly deteriorates. Similar results have been obtained by \cite{LyapunovEquationSolutions}, which examines the numerical accuracy of both algorithms when finding $X^Q(T)$.

\section{Application to an LQG system} \label{s:ApplicationToAnLQGSetUp}

So far we have only considered systems of the form~\eqref{eq:SystemDefinition}, but in LQG systems there are also input and output signals. However, in that case we can always rewrite the system on the form~\eqref{eq:SystemDefinition}. In this section we show how to do this. For more details we refer to \cite{LQBook,MFCBook,DMCSBook,StochasticControlBook}.

First, we consider a system $\dot{\ve{x}}(t) = A\ve{x}(t) + B\ve{u}(t) + \ve{v}(t)$ with input. Its corresponding cost function equals
\begin{equation}
J = \int_0^\infty e^{2\alpha t} (\ve{x}^T(t) Q \ve{x}(t) + \ve{u}^T(t) R \ve{u}(t)) \, dt.
\end{equation}
It is well-known in literature (see for instance \cite{KalmanFilter}) that the optimal control law minimizing $\ex[J]$ is a linear control law $\ve{u}(t) = -F\ve{x}(t)$. If we assume that $Q = Q^T \geq 0$ and $R = R^T > 0$, then the optimal gain matrix $F$ equals
\begin{equation}
F = R^{-1} B^T \hat{X}_\alpha, \label{eq:OptimalGainMatrix}
\end{equation}
with $\hat{X}_\alpha$ the solution to the algebraic Riccati equation
\begin{equation}
A_\alpha^T \hat{X}_\alpha + \hat{X}_\alpha A_\alpha + Q - \hat{X}_\alpha B R^{-1} B^T \hat{X}_\alpha = 0.
\end{equation}
For this optimal gain matrix $F$ (and for any other matrix $F$) the system and cost function can be written as 
\begin{align}
\dot{\ve{x}}(t) & = (A - BF) \ve{x}(t) + \ve{v}(t) = \hat{A} \ve{x}(t) + \ve{v}(t), \\
& \hspace{22pt} J = \int_0^\infty e^{2\alpha t} \ve{x}^T(t) \hat{Q} \ve{x}(t) \, dt,
\end{align}
where we have $\hat{Q} = Q + F^T R F$. This shows that the system is now in our original form~\eqref{eq:SystemDefinition}.

A similar reduction can be performed when we are dealing with a noisy output equation ${\ve{y}(t) = C\ve{x}(t) + \ve{w}(t)}$, where $\ve{w}(t)$ is zero-mean Gaussian white noise with intensity $W$. To deal with this output equation, we take a state estimate $\ve{\hat{x}}(t)$ and update it through
\begin{equation}
\ve{\dot{\hat{x}}}(t) = A \ve{\hat{x}}(t) + B\ve{u}(t) + K(\ve{y}(t) - C\ve{\hat{x}}(t)).
\end{equation}
To minimize the state estimation error $\ve{e}(t) = \ve{\hat{x}}(t) - \ve{x}(t)$, we need to choose the observer gain $K$ equal to
\begin{equation}
K = E C^T W^{-1},
\end{equation}
where $E$ is the solution to
\begin{equation}
AE + EA^T + V - E C^T W^{-1} C E = 0.
\end{equation}
We need this state estimate in a new optimal control law $\ve{u} = -F\ve{\hat{x}}$. This reduces the system equations to
\begin{equation}
\begin{bmatrix}
\ve{\dot{x}} \\
\ve{\dot{\hat{x}}}
\end{bmatrix} = \begin{bmatrix}
A - BF & -BF \\
KC & A - BF - KC
\end{bmatrix} \begin{bmatrix}
\ve{x} \\
\ve{\hat{x}}
\end{bmatrix} + \begin{bmatrix}
\ve{v} \\
K\ve{w}
\end{bmatrix},
\end{equation}
which is again of the form we have seen earlier, albeit with a somewhat larger state vector. Because of this, all the equations that were originally developed for system~\eqref{eq:SystemDefinition} are applicable to LQG systems as well.

\section{Numerical evaluation} \label{s:NumericalEvaluation}

In this section we look at an example of how to apply the derived equations. In literature, researchers almost always use the controller which minimizes the expected value of the cost. This is done irrespective of the variance of the cost. But if the goal is to keep the cost below a certain threshold, then this may not be the best approach.

Consider the two-state system
\begin{equation}\label{eq:SecondExampleSystem}
\ve{\dot{x}} = \begin{bmatrix}
1 & 0 \\
1/20 & 1
\end{bmatrix} \ve{x} + \begin{bmatrix}
1 \\
0
\end{bmatrix} \ve{u} + \ve{v},
\end{equation}
where we will apply $Q = I$, $R = I$ and $\alpha = -0.8$ in the cost function. As control law we use $\ve{u} = -F\ve{x}$. We assume that the state $\ve{x}$ is fully known, and hence only $F$ needs to be chosen. In practice this is often not the case and only a noisy measurement $\ve{y}$ will be available. To solve this, we can apply the theory from Section~\ref{s:ApplicationToAnLQGSetUp} and subsequently choose the observer gain $K$ along with $F$. However, this process is identical to choosing $F$. So for simplicity of presentation, we only consider selecting $F$.

The optimal control matrix follows from~\eqref{eq:OptimalGainMatrix} as $F_{\text{opt}} = \begin{bmatrix}
1.6 & 9.9
\end{bmatrix}$. It minimizes $\ex[J]$ at $\ex[J(F_{\text{opt}})] = 154.4$. However, we can also minimize $\va[J]$ using a basic gradient descent method. This gives the minimum-variance control matrix $F_{\text{mv}} = \begin{bmatrix}
4.4 & 30.0
\end{bmatrix}$ with mean cost $\ex[J(F_{\text{mv}})] = 187.5$. This mean cost is significantly larger than $\ex[J]_{\text{opt}}$, making it seem as if this is a significantly worse control matrix.

However, now suppose that we do not care so much about the mean cost. All we want is to reduce the probability that the cost $J$ is above a certain threshold $\bar{J}$. That is, we aim to minimize $p(J > \bar{J})$ where we use $\bar{J} = 1\thinspace 500$, which is roughly ten times the mean. Using $250 \thinspace 000$ numerical simulations, with \mbox{$T = 20$ s} and \mbox{$dt = 0.01$ s}, we have found that
\begin{align}
p(J(F_{\text{opt}}) > \bar{J}) &\approx 0.091\%, \\
p(J(F_{\text{mv}}) > \bar{J}) &\approx 0.059\%.
\end{align}
Hence the optimal controller has more than half as many threshold-violating cases as the minimum-variance control law, which is a significantly worse result. 

\section{Conclusions} \label{s:ConclusionsAndRecommendations}

In this paper, equations have been derived for the mean and the variance of both the infinite-time cost $J$ and the finite-time cost $J_T$. We have seen a case in which the equations can support controller synthesis by reducing the number of extreme cases that occur.

The infinite-time cost $J$ has a finite value if and only if $A_\alpha$ is stable and $\alpha < 0$. In this case, $\ex[J]$ can be found through Theorem~\ref{th:InfiniteTimeCostMean} and $\va[J]$ through Theorem~\ref{th:InfiniteTimeCostVariance}. The finite-time cost $J_T$ always has a finite value. The theorems needed to find its mean and variance, as well as the requirements for using these theorems, have been summarized in Table~\ref{t:TheoremOverview}. Alternatively, when $T$ is not too large, these two quantities can also be calculated through Theorem~\ref{th:MatrixExponentialsForCostMeanAndVariance} using matrix exponentials for any $A$ and $\alpha$.

\begin{ack}
This research is supported by the Dutch Technology Foundation STW, which is part of the Netherlands Organisation for Scientific Research (NWO), and which is partly funded by the Ministry of Economic Affairs (Project number: 12173, SMART-WIND). The work was also supported by the Swedish research Council (VR) via the project \emph{Probabilistic modeling of dynamical systems} (Contract number: 621-2013-5524).
\end{ack}

\appendix

\section{Evolution of the state} \label{s:StateEvolution}

The way in which the state $\ve{x}(t)$ evolves over time is described by~\eqref{eq:SystemDefinition}. Solving this equation for $\ve{x}(t)$ results in
\begin{equation}
\ve{x}(t) = e^{A t} \ve{x}_0 + \int_0^t e^{A(t-s)} \ve{v}(s) \, ds.\label{eq:StateSolution}
\end{equation}
We use this to derive statistical properties for $\ve{x}(t)$. These properties are well-known (see for instance \cite{DMCSBook}), but they are included to give a good overview of existing theory.

\begin{thm}\label{th:PropertiesOfStateProcess}
When $\ve{x}(t)$ satisfies system~\eqref{eq:SystemDefinition}, with the corresponding assumptions on $\ve{x}(0)$ and $\ve{v}$, then $\ve{x}(t)$ is a Gaussian random variable satisfying
\begin{align}
\ve{\mu}(t) & \hspace{-1pt}\equiv\hspace{-1pt} \ex[\ve{x}(t)] \hspace{-1pt}=\hspace{-1pt} e^{A t} \ve{\mu}_0, \\
\Sigma(t) & \hspace{-1pt}\equiv\hspace{-1pt} \ex[\ve{x}(t) \ve{x}^T(t)] \hspace{-1pt}=\hspace{-1pt} e^{A t} (\Sigma_0 \hspace{-1pt}-\hspace{-1pt} X^V) e^{A^T t} \hspace{-1pt}+\hspace{-1pt} X^V. \label{eq:StateExpectedSquaredValue}
\end{align}
\end{thm}
\begin{pf}
Because $\ve{x}(t)$ is the sum of Gaussian variables, it will have a Gaussian distribution at all times $t$. From~\eqref{eq:StateSolution}, its mean equals
\begin{equation}
\ve{\mu}(t) \equiv \ex[\ve{x}(t)] = e^{A t} \ex[\ve{x}_0] = e^{A t} \ve{\mu}_0.
\end{equation}
The expected squared value is found similarly through
\begin{align}
\Sigma(t) & = e^{A t} \ex[\ve{x}_0 \ve{x}_0^T] e^{A^T t} \nonumber \\
& \hspace{12pt} + \hspace{-2pt} \int_0^t \hspace{-4pt} \int_0^t \hspace{-4pt} e^{A (t-s_1)} \ex[\ve{v}(s_1) \ve{v}^T(s_2)] e^{A^T (t-s_2)} ds_1 \, ds_2 \nonumber \\
& = e^{A t} \Sigma_0 e^{A^T t} + \int_0^t e^{A (t-s)} V e^{A^T (t-s)} \, ds. \label{eq:SigmaTIntermediateExpression}
\end{align}
(The reduction of $\ex[\ve{v}(s_1) \ve{v}^T(s_2)]$ to $V \delta(s_1 - s_2)$ is formally an application of the It\^o isometry, as explained in~\cite{StochasticDEBook}.) Next, by substituting $s$ by $t - \tau$, we find that
\begin{align}
\Sigma(t) & = e^{A t} \Sigma_0 e^{A^T t} + \int_0^t e^{A \tau} V e^{A^T \tau} \, ds \label{eq:DerivationOfStateVariance} \\
& = e^{A t} \Sigma_0 e^{A^T t} + X^V(t) = e^{A t} (\Sigma_0 \hspace{-1pt}-\hspace{-1pt} X^V) e^{A^T t} \hspace{-1pt}+\hspace{-1pt} X^V\hspace{-1pt}, \nonumber
\end{align}
where in the end we have also applied Theorem~\ref{th:FiniteIntegralToLyapunov}.
\end{pf}

\begin{thm}\label{th:StateProcessDerivative}
The expected squared value $\Sigma(t)$ satisfies
\begin{equation}
\dot{\Sigma}(t) = A\Sigma(t) + \Sigma(t)A^T + V. \label{eq:StateProcessDerivative}
\end{equation}
\end{thm}
\begin{pf}
The derivative of~\eqref{eq:StateExpectedSquaredValue} equals
\begin{align}
\dot{\Sigma}(t) & \hspace{-1pt}=\hspace{-1pt} A \hspace{-1pt}\left(\hspace{-1pt}e^{A t} (\Sigma_0 \hspace{-1pt}-\hspace{-1pt} X^V) e^{A^T t}\hspace{-1pt}\right) \hspace{-1pt}+\hspace{-1pt} \left(\hspace{-1pt}e^{A t} (\Sigma_0 \hspace{-1pt}-\hspace{-1pt} X^V) e^{A^T t}\hspace{-1pt}\right)\hspace{-1pt} A^T \nonumber \\
& \hspace{-1pt}=\hspace{-1pt} A \left(\Sigma(t) \hspace{-1pt}-\hspace{-1pt} X^V\right) \hspace{-1pt}+\hspace{-1pt} \left(\Sigma(t) \hspace{-1pt}-\hspace{-1pt} X^V\right) A^T \nonumber \\
& \hspace{-1pt}=\hspace{-1pt} A \Sigma(t) \hspace{-1pt}+\hspace{-1pt} \Sigma(t) A^T \hspace{-1pt}-\hspace{-1pt} \left(AX^V \hspace{-1pt}+\hspace{-1pt} X^V A^T\right).
\end{align}
Applying $AX^V + X^V A^T + V = 0$ completes the proof.
\end{pf}

\begin{thm}\label{th:MultipleTimeStateProcess}
For $t_1 < t_2$ we have
\begin{equation}
\Sigma(t_1,t_2) = e^{A t_1} (\Sigma_0 - X^V) e^{A^T t_2} + X^V e^{A^T (t_2 - t_1)}.
\end{equation}
Furthermore, $\Sigma(t_1,t_2) = \Sigma(t_2,t_1)^T$ and $\Sigma(t,t) = \Sigma(t)$.
\end{thm}
\begin{pf}
The proof is identical to that of Theorem~\ref{th:PropertiesOfStateProcess}.
\end{pf}

\section{Properties of Lyapunov equation solutions} \label{s:LyapunovTheorems}

\begin{thm}\label{th:LyapunovEquationSolutionUniqueness}
There is a unique solution for $X^Q$, and identically for $\bar{X}^Q$, if and only if the matrix $A$ is Sylvester.
\end{thm}
\begin{pf}
In literature it is known (see~\cite{SylvesterEquation}) that the Sylvester Equation $AX + XB = Q$ has a unique solution if and only if $A$ and $-B$ do not have a common eigenvalue. Substituting $B = A^T$ directly proves the theorem.
\end{pf}

\begin{thm}\label{th:SymmetricLyapunovSolution}
Assume that $A$ is Sylvester. In this case $X^Q$ is symmetric if and only if $Q$ is symmetric.
\end{thm}
\begin{pf}
If we take the Lyapunov equation $A X^Q + X^Q A^T + Q = 0$ and subtract its transpose, we find that
\begin{equation}
\hspace{-4pt} A \hspace{-1pt}\left(\hspace{-1pt}X^Q \hspace{-2pt}-\hspace{-2pt} (X^Q)^T\right) \hspace{-1pt}+\hspace{-1pt} \left(X^Q \hspace{-2pt}-\hspace{-2pt} (X^Q)^T\right)\hspace{-1pt} A^T \hspace{-1pt}+\hspace{-1pt} (Q \hspace{-2pt}-\hspace{-2pt} Q^T) \hspace{-2pt}=\hspace{-2pt} 0.\hspace{-8pt}
\end{equation}
This equation has a unique solution (Theorem~\ref{th:LyapunovEquationSolutionUniqueness}) directly implying that $Q = Q^T$ if and only if $X^Q = (X^Q)^T$.
\end{pf}

\begin{thm}\label{th:InfiniteIntegralToLyapunov}
Assume that $A$ is stable. Then $A$ is Sylvester and the Lyapunov equation $A X^Q + X^Q A^T + Q = 0$ has a unique solution $X^Q$ which equals
\begin{equation}
X^Q = \int_0^\infty e^{A t} Q e^{A^T t} \, dt. \label{eq:LyapunovIntegral}
\end{equation}
\end{thm}
\begin{pf}
The assumption that $A$ is stable directly implies that $A$ is Sylvester and hence (Theorem~\ref{th:LyapunovEquationSolutionUniqueness}) that $X^Q$ exists and is unique. Now we only need to prove~\eqref{eq:LyapunovIntegral}. Because $A$ is stable, we know that $\lim_{t \rightarrow \infty} e^{At} = 0$. We can hence write $Q$ as
\begin{align}
Q & = -\left[e^{A t} Q e^{A^T t}\right]_0^\infty = -\int_0^\infty \frac{d}{dt} \left(e^{A t} Q e^{A^T t}\right) \, dt \nonumber \\
& = -\int_0^\infty \left(A e^{A t} Q e^{A^T t} + e^{A t} Q e^{A^T t} A^T\right) \, dt \\
& = -A \hspace{-1pt}\left(\int_0^\infty e^{A t} Q e^{A^T t} \, dt\hspace{-1pt}\right) - \left(\int_0^\infty e^{A t} Q e^{A^T t} \, dt\hspace{-1pt}\right)\hspace{-1pt} A^T. \nonumber
\end{align}
The equation above is a Lyapunov equation with the quantity between brackets as its unique solution $X^Q$.
\end{pf}

\begin{thm}\label{th:FiniteIntegralToLyapunov}
When $A$ is Sylvester, $X^Q(t_1,t_2)$ can either be found by solving the Lyapunov equation
\begin{equation}
A X^Q\hspace{-1pt}(t_1,t_2) +\hspace{-1pt} X^Q\hspace{-1pt}(t_1,t_2) A^T \hspace{-2pt}+ e^{A t_1} Q e^{A^T t_1} \hspace{-1pt}- e^{A t_2} Q e^{A^T t_2} \hspace{-2pt}=\hspace{-2pt} 0\label{eq:FiniteTimeLyapunovEquation}
\end{equation}
or by first finding $X^Q$ and then using
\begin{equation}
X^Q(t_1,t_2) = e^{A t_1} X^Q e^{A^T t_1} - e^{A t_2} X^Q e^{A^T t_2}.\label{eq:FiniteTimeLyapunovEquationPartTwo}
\end{equation}
\end{thm}
\begin{pf}
We first prove~\eqref{eq:FiniteTimeLyapunovEquation} through
\begin{align}
& e^{A t_1} Q e^{A^T t_1} - e^{A t_2} Q e^{A^T t_2} = -\left[e^{A t} Q e^{A^T t}\right]_{t_1}^{t_2} \nonumber \\
& \hspace{12pt} = -\int_{t_1}^{t_2} \frac{d}{dt} \left(e^{A t} Q e^{A^T t}\right) \, dt \nonumber \\
& \hspace{12pt} = -A \hspace{-1pt}\left(\hspace{-1pt}\int_{t_1}^{t_2} e^{A t} Q e^{A^T t} \, dt\hspace{-1pt}\right) \hspace{-1pt}-\hspace{-1pt} \left(\hspace{-1pt}\int_{t_1}^{t_2} e^{A t} Q e^{A^T t} \, dt\hspace{-1pt}\right)\hspace{-1pt} A^T \nonumber \\
& \hspace{12pt} = -A X^Q(t_1,t_2) - X^Q(t_1,t_2) A^T.
\end{align}
To prove~\eqref{eq:FiniteTimeLyapunovEquationPartTwo} too, we will use $Q = -AX^Q - X^Q A^T$ and the matrix property $e^{A t}A = Ae^{A t}$ to find that
\begin{align}
& e^{A t_1} Q e^{A^T t_1} \hspace{-2pt}-\hspace{-2pt} e^{A t_2} Q e^{A^T t_2} = -A \Big(\hspace{-1pt}e^{A t_1} X^Q e^{A^T t_1} \hspace{3pt} \\
& \hspace{5pt} \hspace{-2pt}-\hspace{-2pt} e^{A t_2} X^Q e^{A^T t_2}\hspace{-1pt}\Big) \hspace{-2pt}-\hspace{-2pt} \Big(\hspace{-1pt}e^{A t_1} X^Q e^{A^T t_1} \hspace{-2pt}-\hspace{-2pt} e^{A t_2} X^Q e^{A^T t_2}\hspace{-1pt}\Big) A^T. \nonumber
\end{align}
The above expression actually equals~\eqref{eq:FiniteTimeLyapunovEquation}, except that the part between brackets is replaced by $X^Q(t_1,t_2)$. Because $A$ is Sylvester, the expression has a unique solution $X^Q(t_1,t_2)$, which must equal the part between brackets.
\end{pf}

\begin{thm}\label{th:AddingLyapunovSolutions}
Assume that $A$ is Sylvester and that $AC = CA$. For any $Q$ and $V$ we then have
\begin{equation}
X^{CQ + V} = CX^Q + X^V. \label{eq:AddingLyapunovSolutions}
\end{equation}
\end{thm}
\begin{pf}
Per definition, $AX^Q + X^QA^T + Q = 0$ and $AX^V + X^VA^T + V = 0$. Left-multiplying the first expression by $C$ and adding it to the second gives us
\begin{equation}
\hspace{-0pt} A\left(CX^Q \hspace{-2pt}+\hspace{-2pt} X^V\right) \hspace{-2pt}+\hspace{-2pt} \left(CX^Q \hspace{-2pt}+\hspace{-2pt} X^V\right) A^T \hspace{-2pt}+\hspace{-1pt} \left(CQ \hspace{-2pt}+\hspace{-2pt} V\right) \hspace{-2pt}=\hspace{-1pt} 0. \hspace{-2pt}
\end{equation}
This is a Lyapunov equation with $X^{CQ+V}$ as its solution.
\end{pf}

\begin{thm}\label{th:InterchangeLyapunovSolutions}
Assume that $A$ is Sylvester. For matrices $F$ and $G$ satisfying $AF = FA$ and $A^TG = GA^T$, and for any $Q$ and $V$, we have
\begin{equation}
\tr\left(Q F X^V G\right) = \tr\left(\bar{X}^Q F V G\right).
\end{equation}
\end{thm}
\begin{pf}
This is directly proven by
\begin{align}
\tr\left(Q F X^V G\right) & = \tr\left((-A^T \bar{X}^Q - \bar{X}^Q A) F X^V G\right) \nonumber \\
& = \tr\left((-A^T \bar{X}^Q F X^V G - \bar{X}^Q A F X^V G)\right) \nonumber \\
& = \tr\left((-G \bar{X}^Q F X^V A^T - G \bar{X}^Q F A X^V)\right) \nonumber \\
& = \tr\left(G \bar{X}^Q F (-X^V A^T - A X^V)\right) \nonumber \\
& = \tr\left(\bar{X}^Q F V G\right).
\end{align}
\end{pf}

\begin{thm}\label{th:DifferenceBetweenLyapunovSolutions}
Assume that both $A$ and $A_\alpha$ are Sylvester. For $X^Q$, $X_\alpha^Q$, $X_\alpha^{X^Q}$ and $X^{X_\alpha^Q}$ we have
\begin{equation}
X_\alpha^{X^Q} = \frac{X_\alpha^Q - X^Q}{2\alpha} = X^{X_\alpha^Q}. \label{eq:DifferenceBetweenLyapunovSolutions}
\end{equation}
\end{thm}
\begin{pf}
Per definition, we have
\begin{align}
(A + \alpha I) X_\alpha^Q + X_\alpha^Q (A + \alpha I)^T + Q & = 0, \\
A X^Q + X^Q A^T + Q & = 0.
\end{align}
By subtracting the two equations, and by using $A_\alpha = A + \alpha I$, we can get either of two results
\begin{align}
A (X_\alpha^Q \hspace{-1pt}-\hspace{-1pt} X^Q) \hspace{-1pt}+\hspace{-1pt} (X_\alpha^Q \hspace{-1pt}-\hspace{-1pt} X^Q) A^T \hspace{-1pt}+\hspace{-1pt} 2\alpha X_\alpha^Q & \hspace{-1pt}=\hspace{-1pt} 0, \\
A_\alpha (X_\alpha^Q \hspace{-1pt}-\hspace{-1pt} X^Q) \hspace{-1pt}+\hspace{-1pt} (X_\alpha^Q \hspace{-1pt}-\hspace{-1pt} X^Q) A_\alpha^T \hspace{-1pt}+\hspace{-1pt} 2\alpha X^Q & \hspace{-1pt}=\hspace{-1pt} 0.
\end{align}
Next, we divide the above equations by $2\alpha$. The resulting Lyapunov equations have~\eqref{eq:DifferenceBetweenLyapunovSolutions} as their solution.
\end{pf}

\section{Power forms of Gaussian random variables} \label{s:GaussianPowerForms}

\begin{thm}\label{th:ExpectationOfGaussianPowerFour}
Consider a Gaussian random variable $\ve{x}$ with mean $\ve{\mu}$ and expected squared value $\Sigma \equiv \ex[\ve{x}\ve{x}^T]$. For symmetric matrices $P$ and $Q$ we have
\begin{align}
\ex[\ve{x}^T P \ve{x} \ve{x}^T Q \ve{x}] & = \tr(\Sigma P)\tr(\Sigma Q) + 2\tr(\Sigma P \Sigma Q) \nonumber \\
& \hspace{12pt} - 2\ve{\mu}^T P \ve{\mu} \ve{\mu}^T Q \ve{\mu}. \label{eq:ExpectationOfGaussianPowerFour}
\end{align}
\end{thm}
\begin{pf}
We know from \cite{SCEMBook} (Appendix F.3) that, for symmetric $P$ and $Q$, and for a \emph{zero-mean} process $\ve{y} = \ve{x} - \ve{\mu}$ with covariance $Y = \ex[\ve{y}\ve{y}^T] = \Sigma - \ve{\mu}\ve{\mu}^T$, we have
\begin{equation}\label{eq:ExpectationOfPowerFourZeroMean}
\ex[\ve{y}^T P \ve{y} \ve{y}^T Q \ve{y}] = \tr(Y P)\tr(Y Q) + 2\tr(Y P Y Q).
\end{equation}
If we apply this result to the expansion of
\begin{equation}
\ex[\ve{x}^T P \ve{x} \ve{x}^T Q \ve{x}] = \ex[(\ve{y} + \ve{\mu})^T P (\ve{y} + \ve{\mu}) (\ve{y} + \ve{\mu})^T Q (\ve{y} + \ve{\mu})]
\end{equation}
and rewrite the result,~\eqref{eq:ExpectationOfGaussianPowerFour} follows.
\end{pf}

\begin{thm}\label{th:ExpectationOfDifferentGaussianPowerFour}
Consider Gaussian random variables $\ve{x}$ and $\ve{y}$ with joint distribution
\begin{equation}
\begin{bmatrix}
\ve{x} \\
\ve{y}
\end{bmatrix} \sim \mathcal{N}\left(\begin{bmatrix}
\ve{\mu_x} \\
\ve{\mu_y}
\end{bmatrix}, \begin{bmatrix}
K_{xx} & K_{xy} \\
K_{yx} & K_{yy}
\end{bmatrix}\right).
\end{equation}
Also define $\Sigma_{ab} = K_{ab} + \ve{\mu_a}\ve{\mu_b}^T$, where the subscripts $a$ and $b$ can be substituted for $x$ and/or $y$. For symmetric matrices $P$ and $Q$ we now have
\begin{align}
\ex[\ve{x}^T P \ve{x} \ve{y}^T Q \ve{y}] & = \tr(\Sigma_{xx} P)\tr(\Sigma_{yy} Q) + 2\tr(\Sigma_{yx} P \Sigma_{xy} Q) \nonumber \\
& \hspace{24pt} - 2\ve{\mu_x}^T P \ve{\mu_x} \ve{\mu_y}^T Q \ve{\mu_y}.
\end{align}
\end{thm}
\begin{pf}
This follows directly from Theorem~\ref{th:ExpectationOfGaussianPowerFour} with
\begin{equation}
\ve{x'} = \begin{bmatrix}
\ve{x} \\
\ve{y}
\end{bmatrix}, \hspace{8pt} P' = \begin{bmatrix}
P & 0 \\
0 & 0
\end{bmatrix}, \hspace{8pt} Q' = \begin{bmatrix}
0 & 0 \\
0 & Q
\end{bmatrix}.
\end{equation}
\end{pf}

% Include this if you use bibtex and a bib file to produce the bibliography (preferred). The correct style is generated by Elsevier at the time of printing.
\bibliographystyle{plain}        
\bibliography{bibliography}

\end{document}